\documentclass[prb,twocolumn]{revtex4-1}

\usepackage{amssymb}
\usepackage{amsmath}
\usepackage[dvips]{graphicx}

\begin{document}

\title{Asymmetric magnetic dots: A way to control magnetic properties}
\author{N. M. Vargas$^{1,5}$, S. Allende$^{2,5}$, B. Leighton$^{1,5}$, J.
Escrig$^{1,5}$, J. Mej\'{\i}a-L\'{o}pez$^{3,5}$, D. Altbir$^{1,5}$, and Ivan
K. Schuller$^{4}$}
\address{$^1$ Departamento de F\'{\i}sica, Universidad de Santiago de Chile, USACH,
Av. Ecuador 3493, Santiago, Chile} 
\address{$^2$ Departamento de F\'{\i}sica, FCFM, Universidad de Chile, Casilla
487-3, Santiago, Chile} 
\address{$^3$ Facultad de F\'{\i}sica, Pontificia Universidad Cat\'{o}lica de Chile,
Casilla 306, Santiago 22, Chile} 
\address{$^4$ Physics Department and Center for Advanced Nanoscience (CAN), University of California - San Diego, La Jolla,
California 92093-0319} 
\address{$^5$ Center for the Development of Nanoscience and
Nanotechnology, CEDENNA, Av. Ecuador 3493, Santiago, Chile}

\begin{abstract}
We have used Monte Carlo simulations to investigate the magnetic properties
of asymmetric dots as a function of their geometry. The asymmetry of round
dots is produced by cutting off a fraction of the dot and is characterized
by an asymmetry parameter $\alpha$. This shape asymmetry has interesting
effects on the coercivity ($H_{c}$), remanence ($M_{r}$), and barrier for
vortex and C- state formation. The dependences of $H_{c}$ and $M_{r}$ are non monotonic as a function of $\alpha$ with a well defined minima in these parameters. The vortex enters
the most asymmetric part and exits through the symmetric portion of the dot.
With increasing $\alpha $ the vortex formation starts with a C-state which
persists for longer fields and the barrier for vortex exit diminishes with
increasing asymmetry, thus providing control over the magnetic chirality.
This implies interesting, naively-unexpected, magnetic behavior as a function
of geometry and magnetic field. 

\end{abstract}

\maketitle

\section{Introduction}

Recently much attention was dedicated to the study of regular arrays of
magnetic particles produced by a number of lithographic techniques. Besides
the basic scientific interest in the magnetic properties of these systems,
they may provide the means for the production of new magnetic devices, or as
high-density magnetic recording media\cite{CHAPMAN5321}. The properties
exhibited by these nanostructures are strongly dependent on the geometry,
and therefore understanding the effect of the shape is fundamental for the
development of applications of such materials\cite{ross203}.

The magnetization of nanodots may reverse by one of two possible
mechanisms: vortex nucleation and coherent rotation\cite{igor2009}. Vortex
states are characterized by an in-plane and an out-of-plane magnetization.
The in-plane magnetization is characterized by vortex chirality, defined as
the magnetization direction around the vortex core (clockwise or
counterclockwise). The out-of-plane magnetization is defined by the vortex
core or polarity. In this way, vortices exhibit four different magnetic
states defined by their polarity and chirality.

Methods to control the chirality in single FM layer elements exploit an
asymmetry in the applied field, such as produced by a magnetic force
microscope tip \cite{mironov, jaafar}, a magnetic pulse \cite{gaididei}, a
magnetic field gradient \cite{konoto}, or the magnetization history \cite%
{klaui}.

Alternatively asymmetric disks may provide control over the vortex chirality
with an in-plane magnetic field \cite%
{schneider,taniuchi,vavassori,giesen,wu,dumas}. The effect of geometry on
the vortex nucleation, annihilation and switching field distribution was
explored in 40-nm-thick Ni$_{80}$Fe$_{20}$ disk arrays, with a diameter of
300 nm and different degrees of asymmetry\cite{wu}. These measurements and
micromagnetic simulations showed that the nucleation and annihilation of
vortices vary linearly, while the switching field distribution oscillates
with the ratio of the long/short asymmetry axes. More recently,
studies of arrays of asymmetric Co dots showed that the vortices can be
manipulated to annihilate at particular sites under specific field
orientations and cycling sequences.\cite{dumas}

In this paper, Monte Carlo simulations are used to study the magnetic
configurations and reversal processes of asymmetric dots as a function of
their geometry. The behavior of the chirality, coercive field and remanent
magnetization is studied for non-interacting asymmetric dots as a function
of their aspect ratio. Our results show that the asymmetry determines the
region where vortex nucleation occurs, fixing the chirality of the vortex.

\section{Model}

Our starting point is a uniform circular dot with diameter $d=80$ nm and
height $h=20$ nm. We introduce asymmetries in these dots by cutting specific
sections characterized by a parameter $\alpha =R^{\prime }/R$, as
illustrated in Fig. 1. The field is applied in-plane along the asymmetry direction.

\begin{figure}[tph]
\begin{center}
\includegraphics[width=8cm]{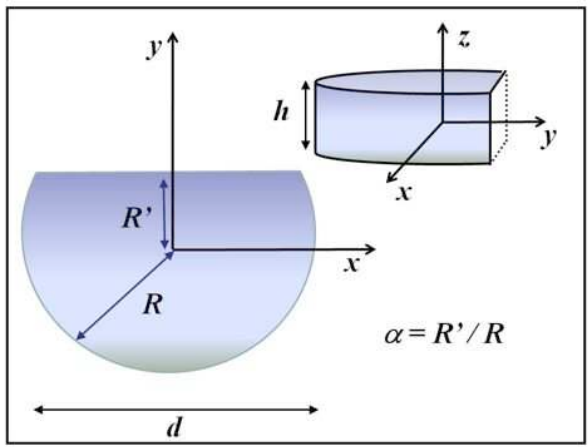}
\end{center}
\caption{(Color online) Geometrical parameters of a nanodot. The white
surface represents the cut surface.}
\end{figure}

A symmetric dot is characterized by $\alpha =1.0$, while a semi circular dot
is given by $\alpha =$ 0.0. To simulate the magnetic properties we
used Monte Carlo simulations, assuming that the interdot distance is large
enough that magnetic interactions are negligible, i.e., each dot behaves
independently~\cite{mejia,CP Li,Grimsditch1}. The internal energy, $E_{tot}$%
, of a single dot with $N$ magnetic moments is given by

\begin{equation}
E_{tot}=\frac{1}{2}\sum\nolimits_{i\neq j}\left( E_{ij}-J_{ij}\hat{\mu}%
_{i}\cdot \hat{\mu}_{j}\right) +E_{H}\,\,,  \label{Etot}
\end{equation}

\noindent where $E_{ij}$ is the dipolar energy given by

\begin{equation}
E_{ij}=\left[ \vec{\mu}_{i}\cdot \vec{\mu}_{j}-3(\vec{\mu}_{i}\cdot \hat{n}%
_{ij})(\vec{\mu}_{j}\cdot \hat{n}_{ij})\right] /r_{ij}^{3}\,,
\label{eq:Edip}
\end{equation}

with \noindent $r_{ij}$ the distance between the magnetic moments $\vec{\mu} _{i}$
and $\vec{\mu}_{j},$ and $\hat{n}_{ij}$ the unit vector along the direction
that connects the two magnetic moments. $J_{ij}$ is the exchange coupling,
which is assumed nonzero only for nearest neighbors and $\hat{\mu}_{i}$ is a
unit vector along the direction of $\vec{\mu}_{i}$. Here $E_{H}=-\sum\nolimits_{i}%
\vec{\mu}_{i}\cdot \vec{H}$ represents the Zeeman energy for a field \ $\vec{%
H}$\ applied along the x direction. As we are interested in polycrystalline
samples, we have not included anisotropy.

Simulation of the magnetic configuration of 10-100 nm structures is not
possible at present with standard computational facilities due to the large
number of magnetic moments within each particle. To avoid this problem we
use a scaling technique developed earlier, \cite{Albuquerque1} for the
calculation of the phase diagram of cylindrical particles. In this approach we define a scaling factor $x$ (0.01-0.001), small enough to reduce the
system to a computationally manageable size while still large enough to
conserve its physical complexity, i.e., for instance the possibility for the
development of a magnetic vortex. With this, physically reasonable results
are obtained, in agreement with micromagnetic calculations, as long as the
exchange constant is rescaled by $J^{\prime }=xJ$, $T^{\prime }=xT$, and $N^{\prime }=Nx^{3\eta }$ with  $\eta \approx 0.55-0.57$. In particular for cylinders, this method allows rescaling geometric parameters
(height, h, and diameter, d, for instance) without loosing physically meaningful
results for the phase diagram and for the general magnetic state of a single nanoparticle. \cite{Vargas1} Thus using this method the geometric parameters are rescaled as $d^{\prime }=dx^{\eta }$ and $h^{\prime }=hx^{\eta }$

For our simulations we use the same parameters used earlier, \cite%
{mejia,mejia2} which produced for symmetric Fe dots good agreement between simulations and experimental measurements. These parameters are
the magnetic moment $|\vec{\mu}_{i}|=\mu =2.2$ $\mu _{B}$, with $\mu _{B}$
the Bohr magneton,  bcc lattice constant $a_{0}=0.28$ nm, and $J=40\ $%
meV. For the dot sizes considered in this paper, $N$ would be larger than $%
10^{7}$, which is computationally unmanageable. Thus we
replace the dot with a smaller one according to the scaling technique
described above. \cite{Albuquerque1,Vargas1,mejia2,mejiap} Correspondingly, we also
scale the exchange interaction by a factor $x\equiv J^{\prime }/J=$0.00245, 
\textit{i.e.}, we replace $J$ with $J^{\prime }=0.098$ meV in the expression
for the total energy. In this case $\eta \approx 0.57$ and $d^{\prime
}=80x^{\eta }=2.68$ nm.

The Monte Carlo simulations are carried out using the Metropolis algorithm
with local dynamics and single-spin flip methods\cite{Binder1}. The new
orientation of the magnetic moment is chosen randomly with a probability $p=%
\mbox{min}[1,exp(-\Delta E/k_{B}T^{\prime })]$, where $\Delta E$ is the
change in energy due to the reorientation of the spin, $k_{B}$ is the
Boltzmann constant, $T^{\prime }$ = $xT$ and $T=10$ K.

The initial state of the system is set up using a random number generator which is used to randomly choose the spin sequence and their individual orientations. A large magnetic field of $H=5.5$ kOe is applied along the [100]
crystallographic direction, labeled as the $x$ axis. This produces a configuration in which the system is saturated and therefore most of the magnetic moments point along this
direction. We define $M_{s}$ as the magnetization at the maximum applied
field ($5.5$ kOe), $M_{r}$ as the remanent magnetization and $H_{c}$ as the
coercivity. Field steps of $\Delta H=0.1$ kOe are used in all calculations,
that is 110 $\Delta H$ values for the complete hysteresis cycle. It is
important to recognize that, due to the non-equilibrium situation, the
number of Monte Carlo steps (MCS) used is a critical issue in the
calculation of the hysteresis loops. Hence, we first study the effect of the
MCS on the coercivity.

Figure 2 illustrates $H_{c}$ for a symmetric dot as a function of MCS. $H_{c}
$ converges asymptotically to 0.47 after 4000 MCS per field value. However,
the effects discussed here are qualitatively similar above MCS $\geq 3500$%
. Therefore we fix the number of Monte Carlo steps for each field at this value,
performing typically $385.000$ Monte Carlo steps per spin for a complete
hysteresis loop. These numbers are independent of the scaling factor, as
discussed in ref. 21. For each calculation six hysteresis loops,
with different random number seeds, are averaged to obtain the results
presented here.

\begin{figure}[tph]
\begin{center}
\includegraphics[width=8cm]{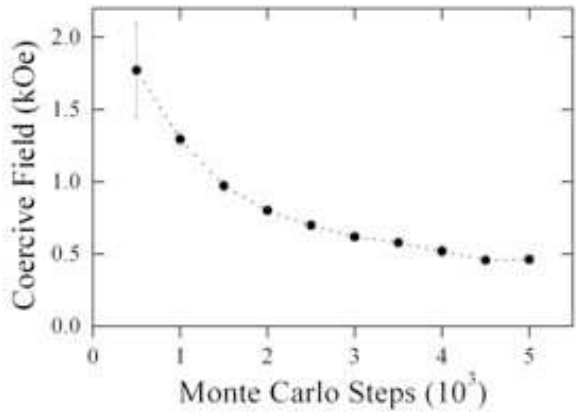}
\end{center}
\caption{Coercivity, $H_{c}$, of a symmetric dot for different numbers of
Monte Carlo Steps, MCS.}
\end{figure}

\section{Results and Discussion}

The main aim of this work is to investigate the effect of the disk shape
asymmetry on the magnetization reversal process. Fig. 3 shows a strong
geometry dependence of the hysteresis curves for different $\alpha $. For $%
0.9<\alpha \leq 1.0$ a neck appears with implies that the reversal occurs by
means of the nucleation and propagation of a vortex. Further decrease of $%
\alpha $ leads to almost square loops and the coercivity and remanence
change as a function of $\alpha $, as shown for $h=20$ and $30$ nm in Figs.
4(a) and 4(b), respectively. Even a small asymmetry, ($\alpha =0.95$),
induces an abrupt decrease of both the coercivity and remanence. However,
further decreases of $\alpha $ produces an increase in the remanence and
coercivity. This is a consequence of the competition between exchange, local
dipolar interactions and geometry. The magnetic moments produced on the new
surface experience a lower exchange interaction facilitating the formation
of a C state. Moreover as expected from the Pole Avoidance Principle~\cite%
{Aharoni}, a C state nucleates to avoid the magnetic pole at the new
surface. The C state, which is the precursor of a vortex, decreases the
coercivity. However, further increase of the asymmetry competes with the
local effects described above, tending to inhibit vortex formation.
Therefore the dependence of the coercivity with $\alpha $ is non monotonic.

\begin{figure}[tph]
\begin{center}
\includegraphics[width=8cm]{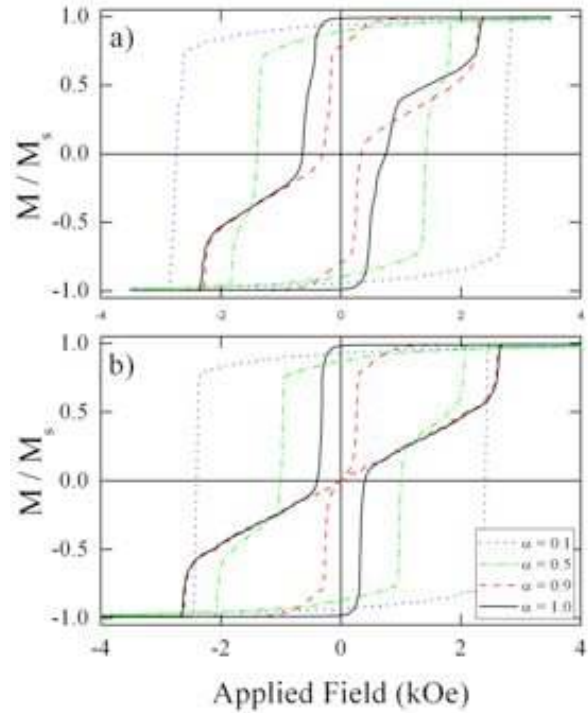}
\end{center}
\caption{(Color online) Hysteresis loops for an asymmetric dot as a
function of $\alpha $ for height $h=20$ nm (a) and $h=30$ nm (b). For $%
\alpha =1.0$ the uniform circular dot has a diameter of $d=80$ nm.}
\end{figure}

\begin{figure}[tph]
\begin{center}
\includegraphics[width=8cm]{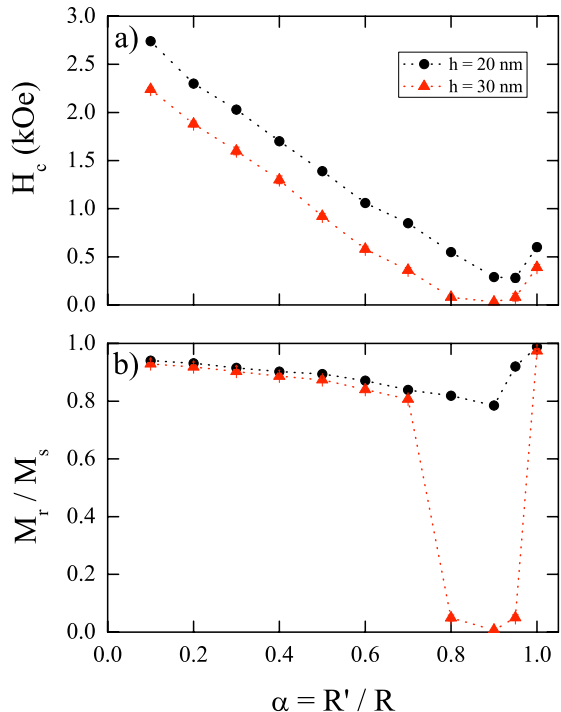}
\end{center}
\caption{(Color online) Coercivity (a) and remanence (b) for asymmetric dots as a function of $\alpha $ for height $h=20$ nm (dots) and $h=30$ nm (triangles). }
\end{figure}

The surprising large quantitative difference between the $h=20$ nm and $h=30$ nm sample in Fig
4(b) is due to the full collapse of the coercivity and the vertical change in
magnetization as a function of the field, as expected from the formation of a
vortex.

We analyze the reversal mechanisms from snapshots of the spin configurations
for different values of $\alpha $ and the applied magnetic field. Figs. 5 (a),
(b) and (c) show snapshots at particular field values for $h=20$ nm and $\alpha
=1.0$, $0.5$ and $0.1$. These snapshots show that all the dots reverse their
magnetization via vortex nucleation and propagation, even the dots with $\alpha
=0.1$, which exhibit almost square hysteresis loops. In symmetric dots,
square loops are a sign of coherent reversal, and the appearance of a neck
indicates that the reversal is driven by a vortex nucleation and propagation%
\cite{mejia}. However for asymmetric dots reversal by vortex nucleation may lead to a
square loop.

\begin{figure}[tph]
\begin{center}
\includegraphics[width=8cm]{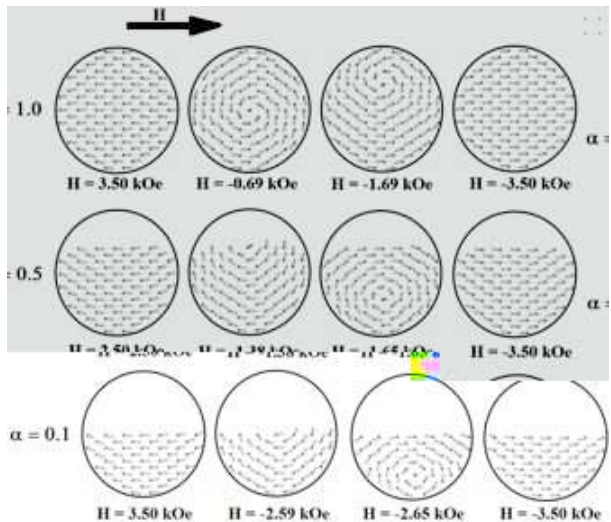}
\end{center}
\caption{Snapshots of the magnetization for a dot with $h=20$ nm at four
different values of $H$. The points depict the position of the magnetic
atoms, while the arrows illustrate the direction of the magnetic moments for $\alpha =1.0$ (a), $0.5$ (b) and 0.1 (c). For (a) the vortex propagates from the $-%
\hat{y}$ to +$\hat{y}$ direction with a clock-wise chirality while for (b) and (c) the vortex propagates from the $+%
\hat{y}$ to -$\hat{y}$ direction with a counterclock-wise chirality.}
\end{figure}

For $\alpha =1.0$ the vortex can nucleate either at the upper or lower
portion of the dot, depending of the seed used in the simulation. For
instance Fig. 5(a) shows the propagation of a vortex which nucleates at the $-y
$ region, while for other seeds nucleation may occur at the opposite region.
However, for $\alpha <1.0$ (Figs. 5(b) and (c)) the nucleation occurs always at
the $+y$ region (the asymmetric part), determining uniquely the chirality, $+z$ (see Fig. 1). This
shows that the asymmetry controls the position of vortex nucleation during reversal whereas the vortex chirality is
determined by the external magnetic field direction. In high magnetic
fields, all spins are aligned along the applied field. For asymmetric dots the reversal proceeds as follows. As the field is reduced at a particular negative field a vortex nucleates with a counterclock-wise direction as viewed from the top (Fig. 1). The reversal starting from saturation in a negative field proceeds in the opposite way. A qualitatively similar behavior occurs for $h=30$ nm. These results are in good agreement
with our independent OOMMF simulations \cite{oommf}.

Finally, we investigate the shape of the vortex as a function of the dot
geometry. To characterize the vortex we define $\beta
=((\sum\nolimits_{i}\mu _{ix})^{2}+(\sum\nolimits_{i}\mu
_{iy})^{2})/M_{s}^{2}$, where $\mu _{ix}$ and $\mu _{iy}$ are the $x$ and $y$
components of individual magnetic moments, and $i$ ranges over all dots. In this way, $\beta =0$ represents a perfect vortex,
while deviations from this state are represented by $\beta \neq 0$. In
particular, $\beta =1$ represents a fully saturated ferromagnetic state. As $%
\beta $ varies from $1$ to $0$, a loop appears and its width represents the
stability of the vortex. Figure 6 shows $\beta $ for different $\alpha $.

In all cases the vortex is almost perfect, $\beta \approx 0$, at some value of the
external field. At this value the magnetization is zero and therefore
represents the coercive field. This value varies with the geometry of the
dot, but also the geometry influences the stability of the C state and the
vortex. In particular, for $\alpha =1.0$ the magnetization reverses by
vortex nucleation at a low field value, $-0.3$ kOe. The abrupt transition of 
$\beta $ from 1 to zero is a consequence of the fast propagation of the
vortex to the center of the dot. This state is very stable, as shown by the $%
-2.2$ kOe field required for vortex annihilation. This feature is
represented in Fig. 6(a) by the continuum transition of $\beta $ from 0 to
approximately 0.3. For $\alpha =0.9$ nucleation of a C-state occurs first at 
$+0.9$ kOe, represented in Fig. 6(b) by the decrease of $\beta $ from 1 to
approximately 0.6. Then, a vortex nucleates at $-0.1$ kOe, which annihilates
at $-1.9$ kOe, after which a C-state appears again, ending the magnetization
reversal. This figure together with the snapshots in Fig. 5 confirms that a
small cut is required for the creation and stabilization of a C-state. The
results for $\alpha =0.5$ are qualitatively similar to those for $\alpha =0.9
$, although the C-state is more stable, as evidenced by the slow decrease of 
$\beta $ from 1 to approximately 0.5, until the vortex appears. During this
reversal the vortex nucleates at $-1.3$ kOe and annihilates at $-1.7$ kOe.
Finally, for $\alpha =0.1$ the C-state is even more stable. The vortex
nucleates at $-2.4$ kOe and annihilates at $-2.6$ kOe. Thus, the
coercivities increase, the C states become more stable, and the vortices
become less stable with decreasing $\alpha $. Interestingly, the degree to
which the $M_{r}$ decreases depends critically on the height of the dot.

\begin{figure}[tph]
\begin{center}
\includegraphics[width=8cm]{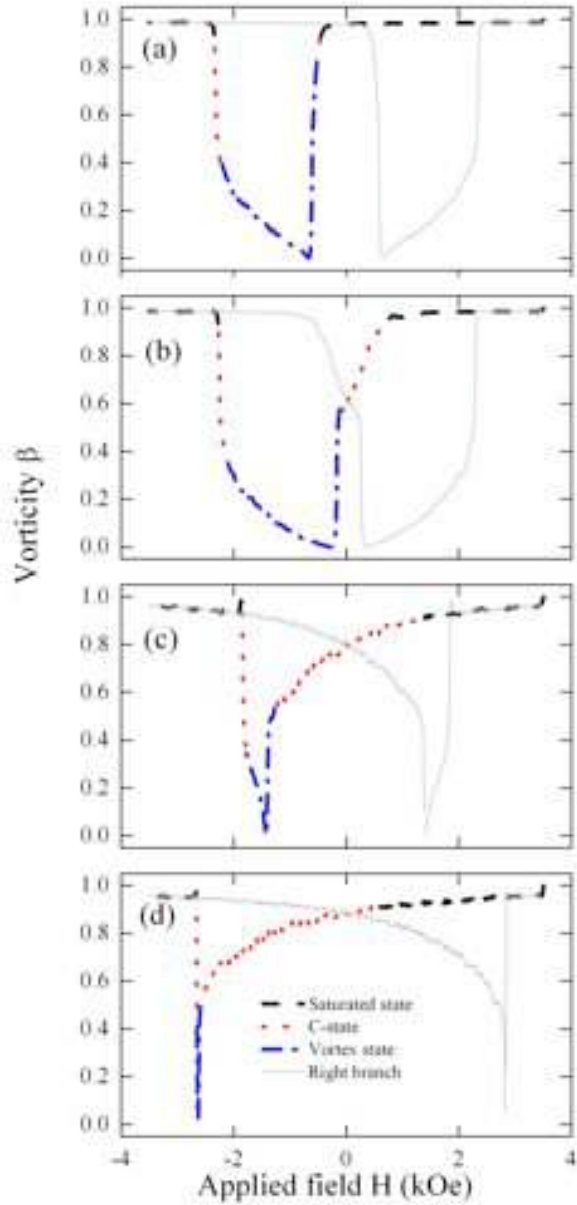}
\end{center}
\caption{(Color online) Vorticity $\beta $ of the left (thick lines) and
right (thin lines) branches of the hysteresis curves as a function of the
external magnetic field, for a) $\alpha =1.0,$ b) $\alpha =0.9,$ c) $\alpha
=0.5$ and d) $\alpha =0.1$.}
\end{figure}

\section{Conclusions}

The results presented above show that the asymmetry determines the region of
vortex nucleation and its chirality during magnetization reversal. The
coercivity, remanence, and vortex stability are strongly affected by the
asymmetry, with a non-monotonic behavior as a function of $\alpha $. These
results are in agreement with previous experimental evidence \cite{wu,dumas}
which explored the relation between asymmetry and chirality. Moreover, it is
showed that all the dots reverse their magnetization via vortex nucleation
and propagation, even dots with $\alpha <0.5$, which exhibit almost square
hysteresis loops. Therefore asymmetry can be used to tailor the magnetic
properties of nanostructured magnetic particles for specific applications.

\section{Acknowledgments}

We acknowledge support from FONDECYT under grants 1080300, 1100365, 3090047
and 11070010, the Millennium Science Nucleus \textquotedblleft Basic and
Applied Magnetism\textquotedblright\ P06-022-F, Financiamiento Basal para
Centros Cient\'{\i}ficos y Tecnol\'{o}gicos de Excelencia, under project FB0807. Research at UCSD was supported by AFOSR
and DOE.

\end{document}